\documentclass[12pt,a4paper]{article}
\usepackage{cite}
\usepackage{url}
\usepackage{amsfonts}
\usepackage{amsmath}
\usepackage{amssymb}
\usepackage{graphicx}
\usepackage{epsfig}
\usepackage{lineno}
\usepackage[usenames]{color}
\usepackage{hyperref}
\usepackage{subfigure}
\usepackage[normalem]{ulem} 
\usepackage{color}
\usepackage{units}
\usepackage{authblk}
\newcommand{\snn}         {\ensuremath{\sqrt{s_{\scriptscriptstyle{{\rm NN}}}}}}

\newcommand{\GeVc}        {GeV/$c$}

\newcommand{\dirg}        {\ensuremath{{\rm \gamma, dir}}}
\newcommand{\decg}        {\ensuremath{{\rm \gamma, dec}}}
\newcommand{\incg}        {\ensuremath{{\rm \gamma, inc}}}
\newcommand{\totg}        {\ensuremath{{\rm \gamma, tot}}}
\newcommand{\bkgg}        {\ensuremath{{\rm \gamma, bkg}}}
\newcommand{\Ndir}        {\ensuremath{N^{\dirg}}}
\newcommand{\Ninc}        {\ensuremath{N^{\incg}}}
\newcommand{\Ndec}        {\ensuremath{N^{\decg}}}
\newcommand{\Ntot}        {\ensuremath{N^{\totg}}}
\newcommand{\Nbkg}        {\ensuremath{N^{\bkgg}}}

\newcommand{\cbkgi}       {\ensuremath{c^{\bkgg, i}}}
\newcommand{\Npiz}        {\ensuremath{N^{\rm{\pi^{0}}}}}
\newcommand{\Rg}          {\ensuremath{R_{\rm \gamma}}}
\newcommand{\vdirg}       {\ensuremath{v_{2}^{\dirg}}}
\newcommand{\vdecg}       {\ensuremath{v_{2}^{\decg}}}
\newcommand{\vincg}       {\ensuremath{v_{2}^{\incg}}}
\newcommand{\vpizero}     {\ensuremath{v_{2}^{\pi^{0}}}}
\newcommand{\vtotg}       {\ensuremath{v_{2}^{\totg}}}
\newcommand{\vbkgg}       {\ensuremath{v_{2}^{\bkgg}}}
\newcommand{\vbkggi}      {\ensuremath{v_{2}^{\bkgg, i}}}
\newcommand{\pion}        {\ensuremath{\rm \pi}}
\newcommand{\kaon}        {\ensuremath{\rm K}}
\newcommand{\prot}        {\ensuremath{\rm p}}
\newcommand{\neu}         {\ensuremath{\rm n}}
\newcommand{\pT}          {\ensuremath{p_{\rm T}}}

\newcommand{\hrefurl}[1]  {\href{#1}{\url{#1}}}
\newcommand{\Eq}[1]       {Eq.~\ref{#1}}
\newcommand{\Ref}[1]      {Ref.~\cite{#1}}

\newcommand{\Fig}[1]      {Fig.~\ref{#1}}

\newcommand{\Sec}[1]      {Sec.~\ref{#1}}

\newcommand{\lsim}        {\,{\buildrel < \over {_\sim}}\,}
\newcommand{\gsim}        {\,{\buildrel > \over {_\sim}}\,}
\newcommand{\co}[1]       {}

\newif\iflatexdiff
\graphicspath{{./img/}}
\usepackage{lineno,xcolor}
\begin{document}
\title{Impact of residual contamination\\ on inclusive and direct photon flow}
\author[1,2]{F.\ Bock}
\author[1]{C.\ Loizides}
\author[3]{T.\ Peitzmann}
\author[3]{M.\ Sas}
\affil[1]{\small LBNL, Berkeley, CA, 94720, USA}
\affil[2]{\small Physikalisches Institut, Ruprecht-Karls-Universit\"at, Heidelberg, Germany}
\affil[3]{\small Utrecht University/Nikhef, Utrecht, Netherlands}
\def\dver{v1.1}
\date{\today, \dver}               
\maketitle
\begin{abstract}
Direct photon flow is measured by subtracting the contribution of decay photon flow from the measured inclusive photon flow via the double ratio $R_{\rm \gamma}$, which defines the excess of direct over decay photons. 
The inclusive photon sample is affected by a modest contamination from different background sources, which is often ignored in measurements. 
However, due to the sensitivity of the direct photon measurement even a residual contamination may significantly bias the extracted direct photon flow.
In particular, for measurements using photon conversions, which are very powerful at low transverse momentum, these effects can be substantial.
Assuming three different types of correlated background contributions we demonstrate using the Therminator2 event generator that the impact of the contamination on the magnitude of direct photon flow can be on the level of $50\%$, even if the purity of the inclusive photon sample is about $97\%$.
Future measurements should attempt to account for the contamination by measuring the background contributions and subtracting them from the inclusive photon flow. 
\end{abstract}
\section{Introduction}
\label{sec:intro}
Heavy-ion collisions at the Relativistic Heavy Ion Collider (RHIC) and the Large Hadron Collider (LHC) produce a state of matter where quarks and gluons are deconfined. 
This state of hot dense matter, the Quark-Gluon Plasma (QGP) is predicted by numerical solutions of Quantum Chromodynamics~\cite{Bhattacharya:2014ara}. 

One way to study the properties of the QGP is by measuring direct photons. 
Direct photons, i.e.\ all photons excluding those from hadronic decays, are produced during every stage of the collision evolution. 
They can be categorized in two regimes governed by different production mechanisms, which to a large extent coincide with specific transverse momentum ranges~\cite{Kapusta:1991qp}.
Prompt direct photons are produced in hard scatterings of incoming partons, dominating the direct photon spectrum at higher transverse momenta ($\pT\gsim4$~\GeVc). 
Thermal direct photons are emitted during the hot QGP and hadron gas phases and dominate at lower transverse momenta ($\pT\lsim4$~\GeVc). 
Since photons interact only weakly with the strongly coupled medium they provide unique information of the produced system allowing one to deduce the initial temperature of the QGP from calculations. 

Direct photon spectra at low $\pT$ have been measured by the WA98~\cite{Aggarwal:2000th}, PHENIX~\cite{Adare:2008ab,Adare:2009qk,Adare:2014fwh} and ALICE~\cite{Adam:2015ldab} collaborations using direct~(real photon) and indirect~(virtual photon) detection techniques.
In the direct approach one usually measures the ratio of inclusive photons over those from decays, quantified with the double ratio 
\begin{equation}
 \label{rg}
 \Rg = \frac{\Ninc}{\Ndec} = {\left(\frac{{\rm d}\Ninc/{\rm d}\pT}{{\rm d}\Npiz/{\rm d}\pT}\right)}\left/{\left(\frac{{\rm d}\Ndec/{\rm d}\pT}{{\rm d}\Npiz/{\rm d}\pT}\right)_{\rm MC}}\right.\,, 
\end{equation}
in which correlated systematic uncertainties approximately cancel.
The direct photon spectra are simply given by the difference of $\Ninc$ and $\Ndec$, expressed as $\Ndir=\left(1-1/\Rg\right)\Ninc$.
At low $\pT$, an enhancement of direct photons by $10$--$20$\% is observed in central AA collisions, not seen in pp or dAu collisions~\cite{Aggarwal:2000th,Adare:2008ab,Adare:2009qk,Adare:2014fwh,Adam:2015ldab}.

Similarly, an azimuthal anisotropy (a.k.a. elliptic flow) of direct photons can be quantified by subtracting the decay from the inclusive photon flow weighted by their respective abundances, $\Ndir\vdirg=\Ninc\vincg-\Ndec\vdecg$, which can be expressed via $\Rg$ as
\begin{equation}
 \label{v2directform}
 \vdirg = \frac{\Rg \vincg - \vdecg}{\Rg-1}\,.
\end{equation}
Measurements by the PHENIX~\cite{Adare:2011zr,Adare:2015lcd} and ALICE~\cite{Lohner:2012ct} collaborations reported a suprisingly large azimuthal anisotropy of direct photons, comparable to that of hadrons.
These observations suggest that the photon production occurs at later stages of the collision when the collective flow of the system is fully developed, while the temperature and, hence, the corresponding thermal photon rates are already reduced. 
Indeed, it is a challenge for models to simultaneously describe the observed direct photon yields and azimuthal anisotropy at low $\pT$, which is referred to as ``the direct photon puzzle'', and has lead to a large amount of recent theoretical effort to resolve the puzzle~\cite{Shen:2013vja,Shen:2013cca,Chatterjee:2013naa,Linnyk:2013wma,vanHees:2014ida,Monnai:2014kqa,McLerran:2014hza,Campbell:2015jga,McLerran:2015mda,Paquet:2015lta,Holt:2015cda,Vovchenko:2016ijt}.

Experimentally, the measurement of direct photons has also been continuously further scrutinized. 
In particular, there has been significant effort to improve the understanding of the systematic uncertainties by measuring photons with different reconstruction techniques, using calorimeters~\cite{Adare:2008ab,Adare:2009qk,Adam:2015ldab}, photon conversions~\cite{Adam:2015ldab}, and a combination of both~\cite{Adare:2014fwh}. 

Since $\Rg$ is close to 1, and the inclusive and decay photon flow are similar in magnitude, the extracted direct photon flow is extremely sensitive to small corrections to $\vincg$.
In order to access the sensitivity of the $\vdirg$ measurement to a remaining contamination from a hadronic background, we use a parameterization of the ALICE preliminary data~\cite{Lohner:2012ct} and the event generator Therminator2~\cite{Chojnacki:2011hb} to model the residual background.
We demonstrate that even neglecting a small contamination in the measured $\vincg$ can lead to significant changes of $\vdirg$.
The article is devided into the following sections:
In \Sec{sec:imp} we briefly recall the measurement of $\vincg$, with particular emphasis on the purity.
In \Sec{sec:bgk} we discuss a realistic model for simulating a residual contamination in $\vincg$.
In \Sec{sec:res} we present and discuss the effects of subtracting various contributions of the residual background on $\vincg$ and $\vdirg$.
In \Sec{sec:sum} we conclude with a short summary.

\section{Inclusive photon flow measurement and purity}
\label{sec:imp}
The elliptic flow of the inclusive photon sample has been measured by reconstructing photons in calorimeters~\cite{Adare:2011zr} and via the photon conversion~\cite{Adare:2015lcd,Lohner:2012ct} method.
In both cases, one attempts to have a clean photon sample free from major contributions of background sources, and thus the photon purity has been maximized. 
For photons reconstructed with a calorimeter one expects that the impurities arise from single misidentified particles~(e$^{\pm}$, $\pion^{\pm,0}$, $\neu$, \ldots). 
For photons reconstructed from conversions into electron-positron pairs one expects a combinatorial background from misidentified pairs~(e$^{-}$+e$^{+}$, $\pion^{\pm}$+e$^{\mp}$, $\kaon^{\pm}$+$e^{\mp}$, $\prot^{\pm}$+e$^{\mp}$, \ldots).
These background sources potentially carry elliptic flow and thus may affect the measured $\vincg$. 

To be able to correct for the residual contamination one needs to know the $v_2$ of the various background sources. 
The $v_2$ for different contributions is additive, so one generally has $\Ntot\vtotg = \Ninc\vincg + \Nbkg\vbkgg$.
Experimentally, one can not obtain an inclusive photon sample with $100$\% purity, i.e.\ $0$\% contamination~($c$), which implies that $\vtotg \neq \vincg$. 
However, if $c=\Nbkg/\left(\Ninc+\Nbkg\right)$ and $\vbkgg$ are known, the $\vincg$ can be corrected using
\begin{equation}
\label{v2correction}
 \vincg = \frac{\vtotg-\sum_{i=0}^{n}{\vbkggi\cbkgi}}{1- \sum_{i=0}^{n}{\cbkgi}}\,,
\end{equation}
where $i$ denotes all possible background sources~($1\le i\le n$), which have to be estimated separately. 
If $c_{i}=0$, there is no correction, as expected.
However, if at least one $c_{i}>0$, there will be a correction depending on the purity and the strength of the respective $\vbkggi$.
For photons reconstructed via conversion electrons, typical values for $c$ at low $\pT$ are about $5$\%~\cite{Lohner:2013blo} and $1$\%~\cite{Adare:2014fwh}. 
In the case of the PHENIX calorimeter-based measurement~\cite{Adare:2011zr} the contamination of $20$\% from hadrons at low $\pT$ was already subtracted using \Eq{v2correction}, with $\vbkgg$ replaced by the measured hadron $v_2$, and an uncertainy of about $2$\% was assigned.
It is important to realize that due to the small value of $\Rg$ even a small change in $\vincg$ has a large effect on $\vdirg$, since it gets amplified by a factor of $5$--$10$, as can be seen from \Eq{v2directform}.

\begin{figure}[t]
  \centering
  \includegraphics[width=0.95\linewidth]{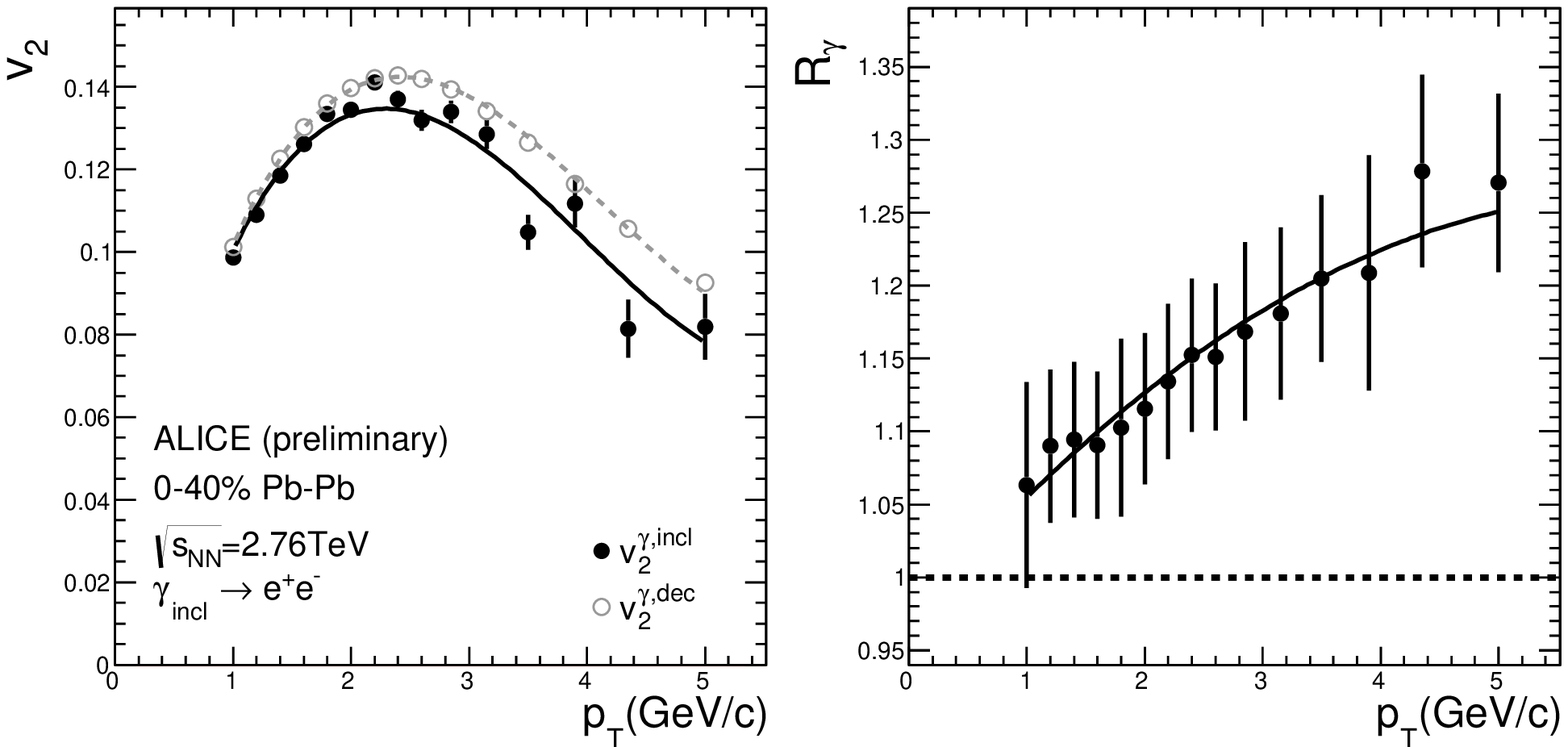}
  \caption{ALICE preliminary results for $\vincg$ and $\vdecg$~(left panel) and $\Rg$~(right panel) measured using the photon conversion method in $0$--$40$\% PbPb collisions at $\snn=2.76$ TeV. The data are parametrized by a 3rd order polynomial for $v_2$ and 2nd order polynomial for $\Rg$ to reduce the effect of fluctuations. The data points are from figure 4 in \Ref{Lohner:2012ct} and 5 in \Ref{Wilde:2012wc}, respectively.}
  \label{fig:ALICEInput}
\end{figure}

In the following, we mostly concentrate on studies of measurements using photon conversion for two reasons:
i) Photon conversion measurements have a large impact at low $\pT$, where the signals related to possible thermal production are expected to be strongest.
ii) The effect of contamination is enhanced in a pair measurement compared to that of single particles.
To study the impact of the inclusive photon impurities on direct photon flow, the ALICE preliminary results~\cite{Lohner:2012ct,Wilde:2012wc} are used as input to the calculation together with a model for $\vbkgg$ discussed further below.
The preliminary results for $\vincg$ and $\vdecg$ and $\Rg$, shown in \Fig{fig:ALICEInput}, have been measured using the photon conversion method in Pb-Pb collisions at $\snn=2.76$ TeV for $0$--$40$\% centrality.
In order to reduce the effect of fluctuations in the subsequent calculation, the data are parametrized by a 3rd order polynomial for $\vincg$ and $\vdecg$ and 2nd order polynomial for $\Rg$, respectively.
The contamination of the inclusive photon sample at low $\pT$ is about 5\%~\cite{Lohner:2013blo}, strongly depending on $\pT$ and centrality. 

\section{Model of the inclusive photon flow background}
\label{sec:bgk}
In order to illustrate the effect that the purity correction from \Eq{v2correction} may have on the inclusive and direct photon $v_2$, we construct a toy model for $\vbkgg$.
We assume that $\vbkgg$ gets a contribution from charged pion flow, since in the photon conversion method the electrons are selected using ${\rm d}E/{\rm d}x$ information of the TPC.
In particular at low $\pT$, the selection regions for electrons and pions overlap, and thus there are pions being misidentified as electrons. 
Since pions carry a $v_2$, a fake photon reconstructed from a $\pion^{\pm}$+e$^{\mp}$ pair will do so as well. 
The same argument holds for kaons and protons and their combinations.
In addition, one also would expect a non-trivial effect from background e$^{+}$+e$^{-}$ pairs. 
For the ALICE preliminary measurement the largest contribution to the combinatorial background are the e$^{+}$+e$^{-}$, closely followed by combinations of $\pion^{\pm}$+e$^{\mp}$. 
These contributions only show a mild transverse momentum dependence, while others contribute mainly below $2$~\GeVc, like $\prot$+$e^{\pm}$, $\kaon^{\pm}$+e$^{\mp}$, $\prot$+$\pion^{\pm}$, due to the crossing points of the respective particles with the electron ${\rm d}E/{\rm d}x$ expectation. 
At high transverse momenta on the other hand charged pions are misidentified as electrons more often and thus the $\pion^+$+$\pion^-$ contributions plays a larger role~\cite{Wilde:2015thesis}.
In reality, these contributions to the inclusive photon sample and their $v_2$ should be measured in data~(or estimated using detector simulations) and subtracted. 

\begin{figure}[t!]
\includegraphics[width=0.329\linewidth]{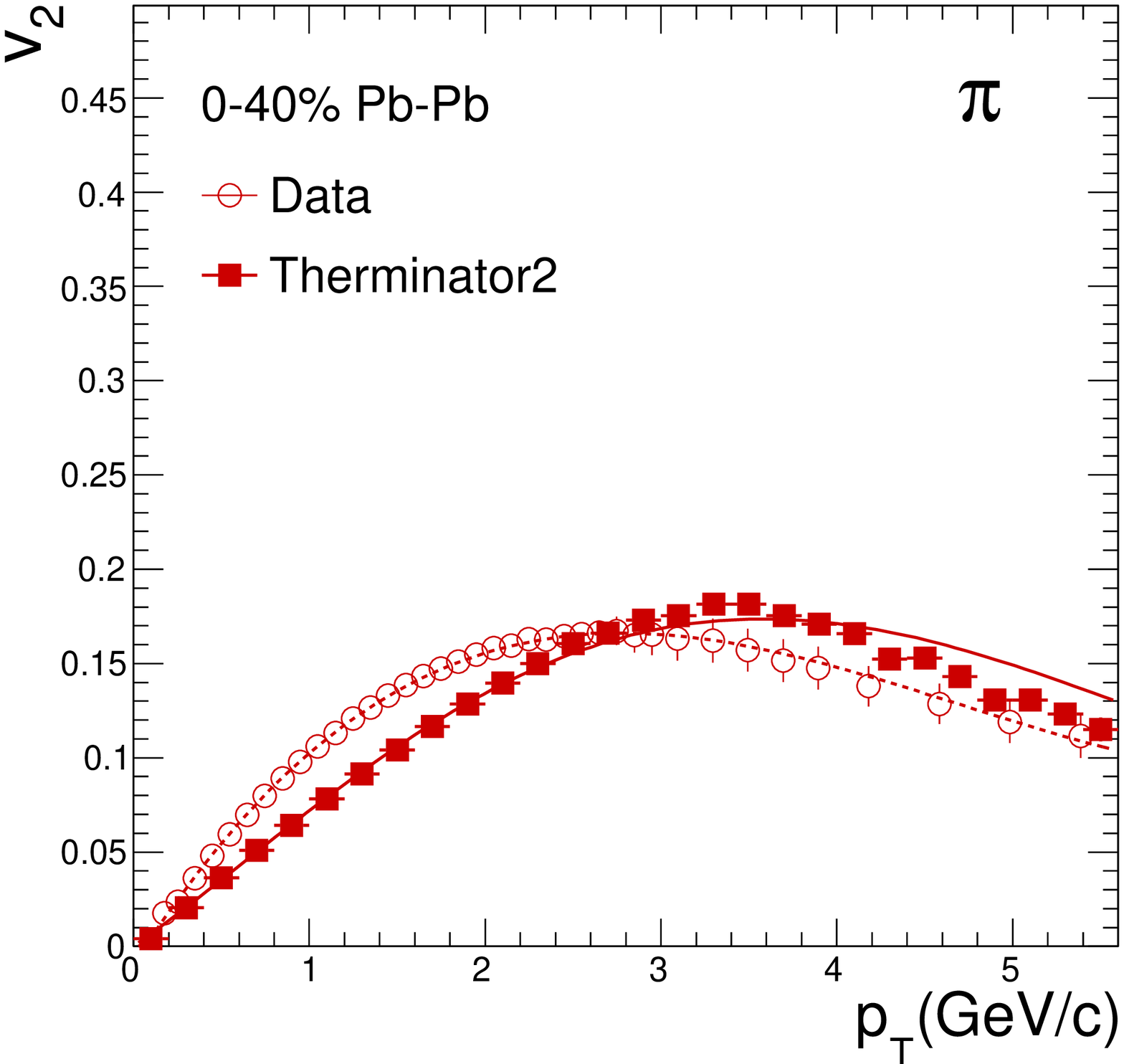}
\includegraphics[width=0.329\linewidth]{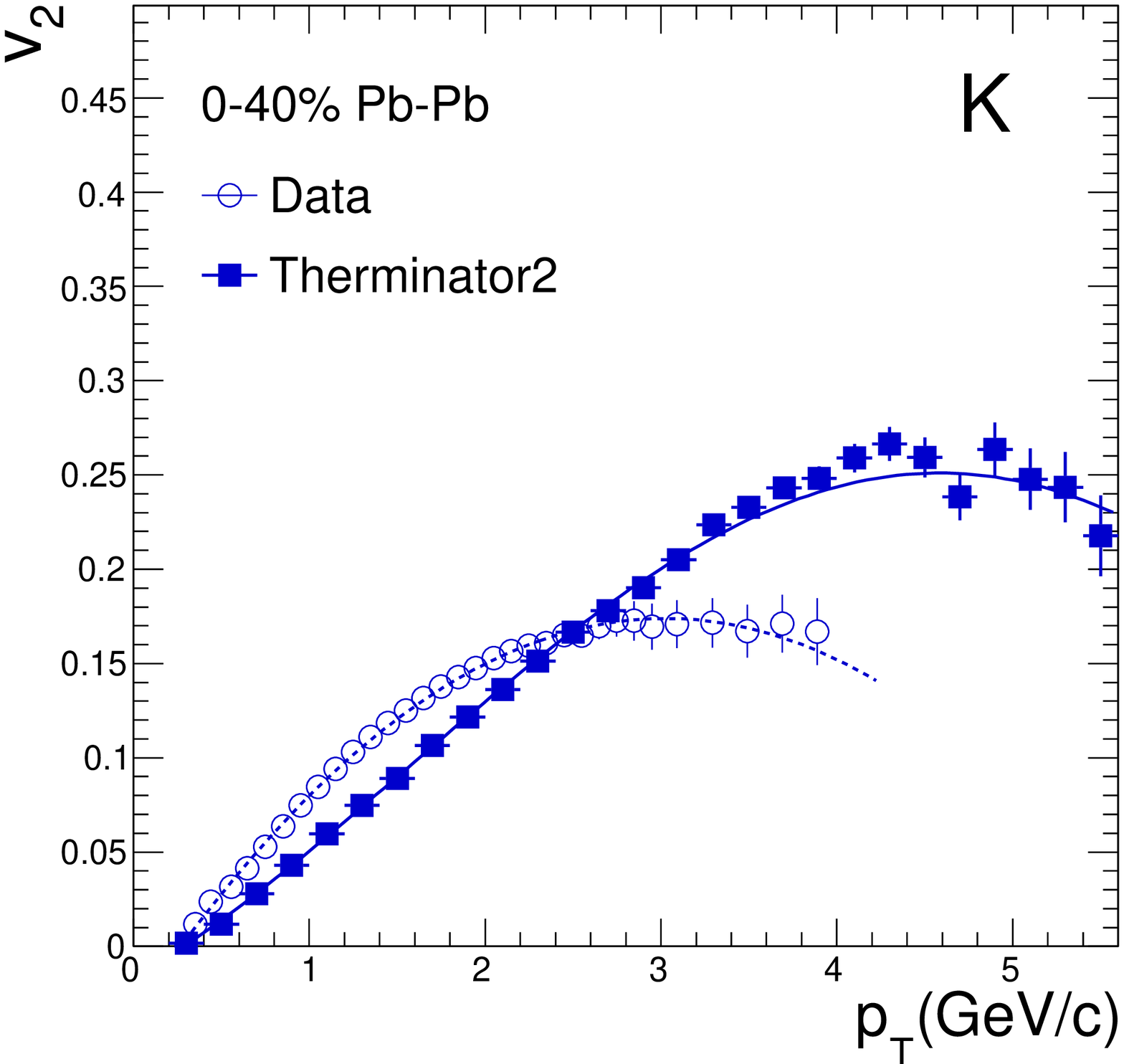}
\includegraphics[width=0.329\linewidth]{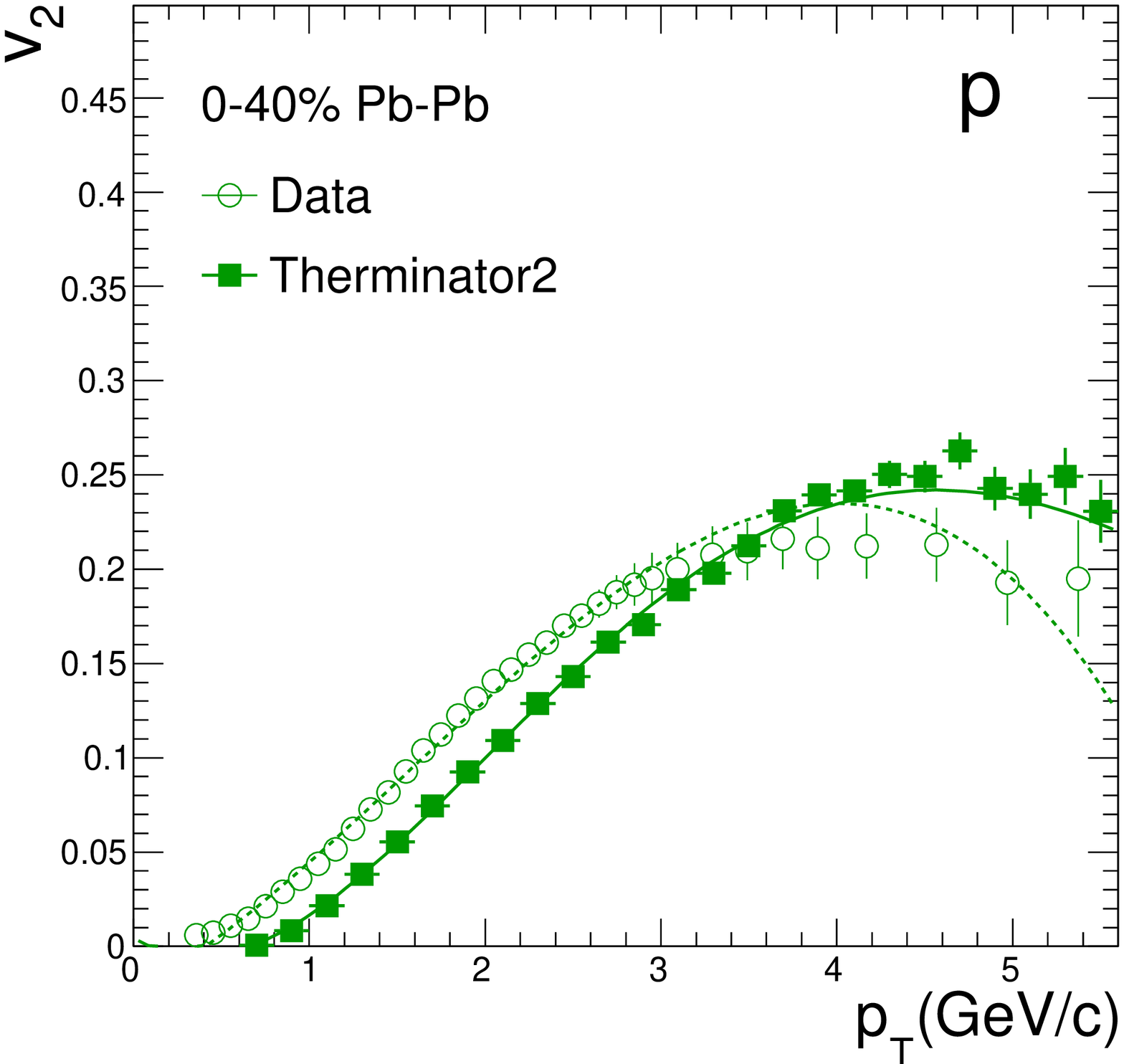}
\caption{Single particle $v_2$ as a function of $\pT$ for $\pion$, $\kaon$ and $\prot$ calculated using Therminator2~\cite{Chojnacki:2011hb} and compared to the measured data~\cite{Abelev:2014pua} for $0$--$40$\% PbPb collisions at $\snn=2.76$~TeV. Fits to both sets of results are also included. For the experimental data the fit uses a 3rd order polynomial~(dashed line) and for Therminator2 a 5th order polynomial~(continous line).}
\label{fig:v2singleflow}
\end{figure}
\begin{figure}[t!]
\includegraphics[width=0.329\linewidth]{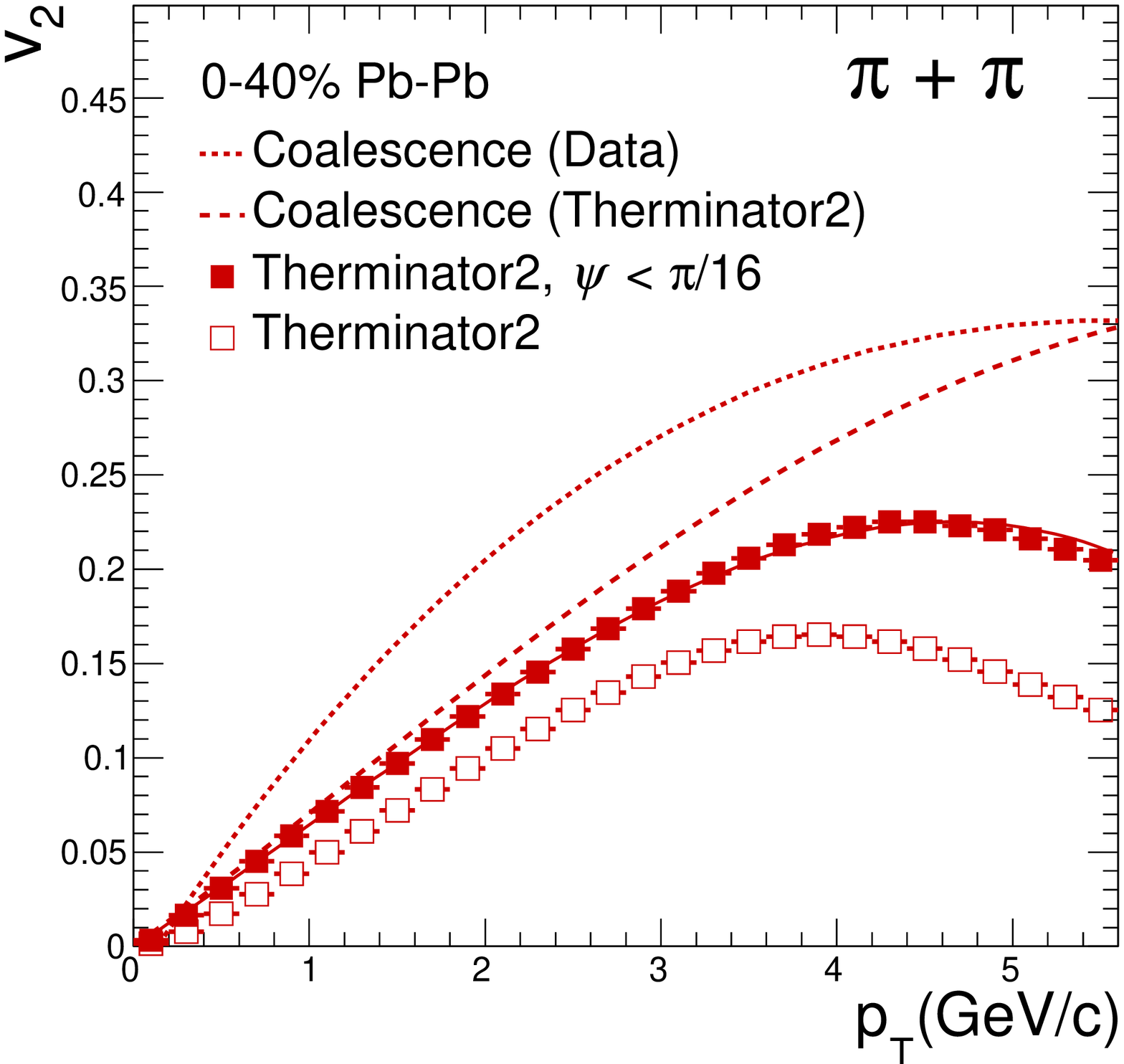}
\includegraphics[width=0.329\linewidth]{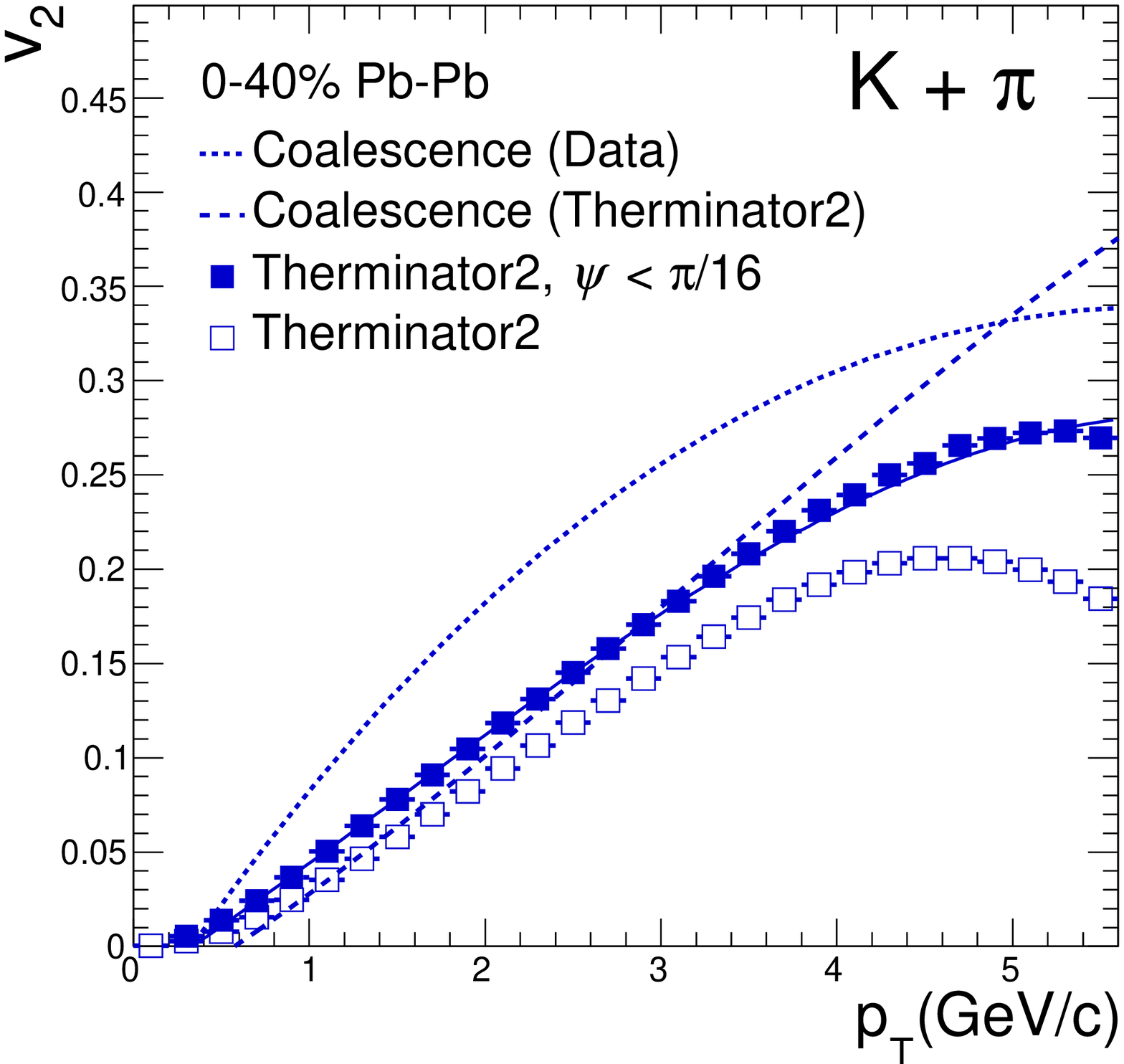}
\includegraphics[width=0.329\linewidth]{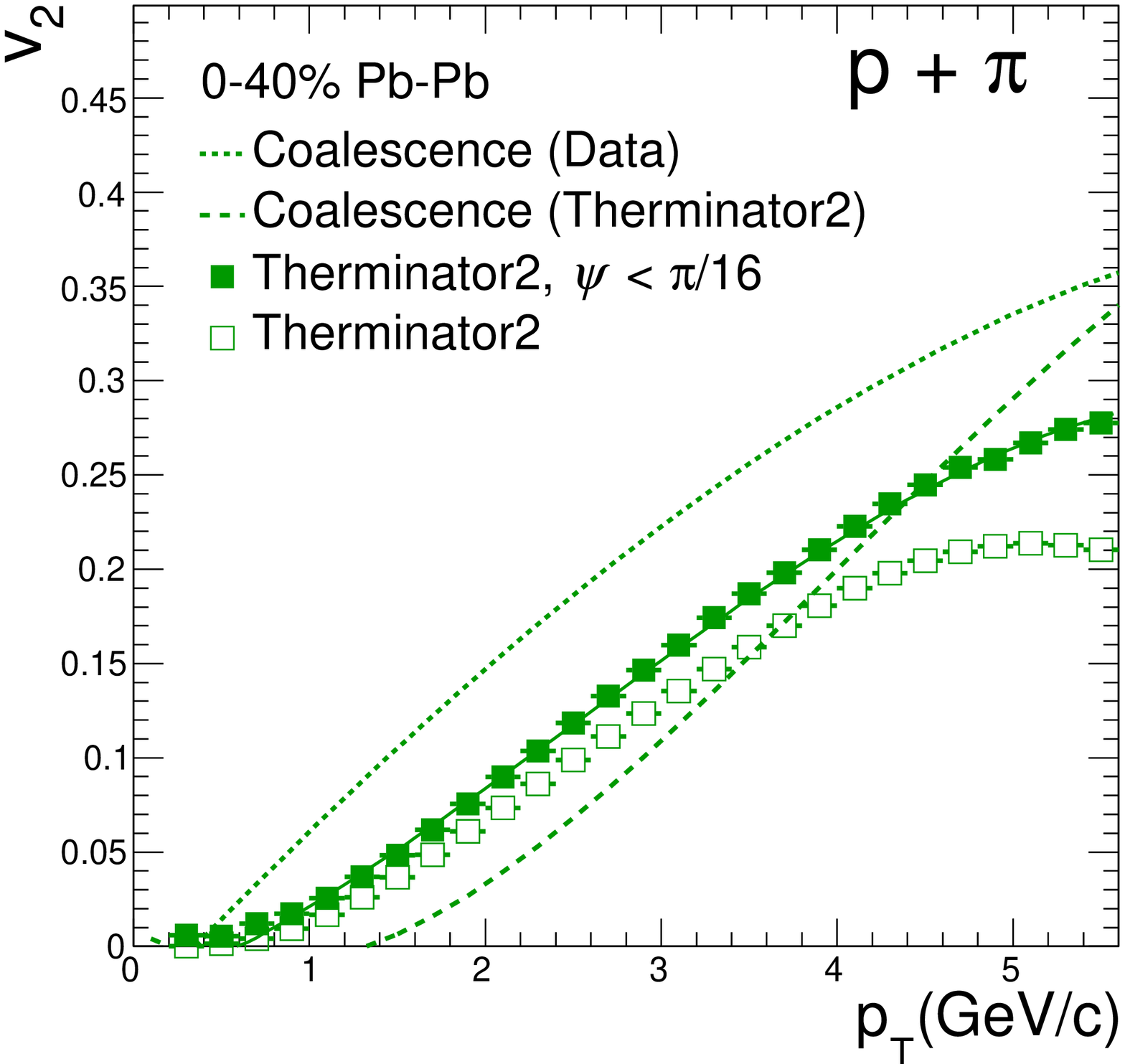}
\caption{Pair $v_2$ as a function of $\pT$ simulated with Therminator2 (symbols) and derived from single-particle $v_2$ using a simple coalescence model. The pair $v_2$ is calculated for $\pion + \pion$, $\kaon + \pion$ and $\prot + \pion$ systems and parametrized by a 3rd order polynomial. The Therminator2 calculations are performed either without any opening angle cut (open symbols) or with a cut of $\pi / 16$ (filled symbols). The coalescence type estimates use as input a parametrization of either the data (dotted line) or of the single particle results from Therminator2 (dashed line).}
\label{fig:v2backgroundflow}
\end{figure}

Instead, here, we simulate the possible pair background using the event generator Therminator2~\cite{Chojnacki:2011hb} in Pb-Pb collisions $\snn=5.02$ TeV for $0$--$40$\% centrality, employing (2+1)-dimensional boost-invariant hydrodynamics. 
The reaction plane is known from the generator output, and the particle $v_2$ can be calculated using the 3-momentum vectors.
The single-particle $v_2$ results for $\pion$, $\kaon$ and $\prot$ are shown in \Fig{fig:v2singleflow} and compared to the measured data~\cite{Abelev:2014pua} for $0$--$40$\% PbPb collisions at $\snn=2.76$~TeV. 
The model reproduces the data reasonably well. 
Only for kaons at larger $\pT$ substantial differences are observed.

Since without detector material the simulation does not contain converted electrons, we can only calculate the contributions from the pair $v_2$ for $\pion+\pion$, $\kaon+\pion$ and $\prot+\pion$ systems by summing up the 3-vectors of the two particles. 
It can be assumed, that at least a fraction of the initial $\vpizero$, from which most electron will originate, will be carried by the electrons. 
Thus in order to obtain a first estimate on the possible effects, we take the $\pion+\pion$ contribution as a reasonable approximation for the $e^{+}$+$e^{-}$ contribution as well, though it will most likely underestimate the strength of the $v_{2}$ at low $\pT$.
The pair $v_2$ results from the Therminator2 simulations are shown in \Fig{fig:v2backgroundflow}, for pairs without any requirement on the opening angle $\psi$ and with a cut of $\psi < \pi/16$. 
The latter cut is applied to mimic the conversion photon selection.
As expected, a significant pair $v_2$ develops for both cases, and the values increase for the smaller opening angle.
The effect observed resembles that caused by the coalescence mechanism~\cite{Voloshin:2002wa,Molnar:2003ff}. 
Stricter cuts on the opening angle select two particles, which are closer in phase space. 
In the construction of the pair the $\pT$ of the single particles are combined.
For small $\psi$ this is equivalent to the sum. 
As a result, the pairs carry a stronger correlation at a higher $\pT$, similar to coalescence models. 

The behaviour suggests that one might use the simple analytical scaling predicted by na\"ive coalescence models to calculate the pair $v_2$. 
Following this idea, we calculate the pair $v_2$ estimated from measured $\pi$, $\kaon$ and $\prot$ $v_2$ data~\cite{Abelev:2014pua} as $v_2^{a+b}(\pT)=v^{a}_{2}(\pT/2)+v_2^{b}(\pT/2)$ for particle species $a$ and $b$. 
In addition, we have also applied the same summation to the single particle $v_2$ generated by Therminator2. 
For this purpose, the $v_2$ results were parameterized by a 3rd order polynomial for the experimental data and by a 5th order polynomial for Therminator2, as shown in \Fig{fig:v2singleflow}. 

The results for pair $v_2$ obtained by the two coalescence-like estimates are also shown in \Fig{fig:v2backgroundflow}.
The estimates of the pair $v_2$ from coalescence are found to be qualitatively similar, with the pair $v_2$ obtained using the parameterization from the data having a steeper increase with $\pT$ than the ones from Therminator2. 
For the purpose of this study the Therminator2 model provides a reasonable description, despite the fact that the model does not perfectly describe the data.
For the following analysis, we use the parameterization of  pair $v_2$ coefficients from Therminator2 with an opening angle cut of $\psi < \pi/16$,
when applying the contamination correction in \Eq{v2correction}.

\begin{figure}[t!]
\centering
\includegraphics[width=0.43\linewidth]{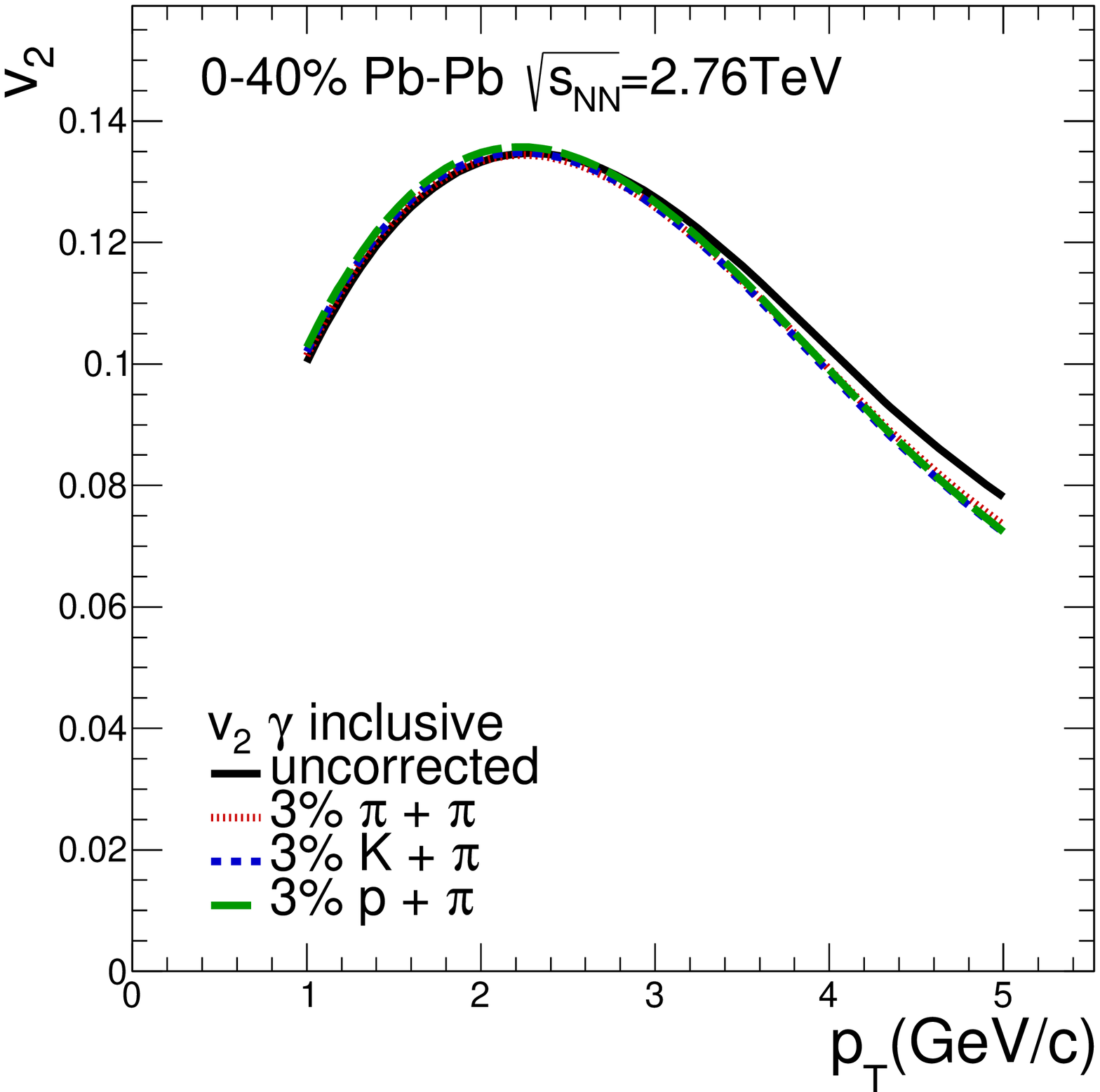}\hspace{0.5cm}
\includegraphics[width=0.43\linewidth]{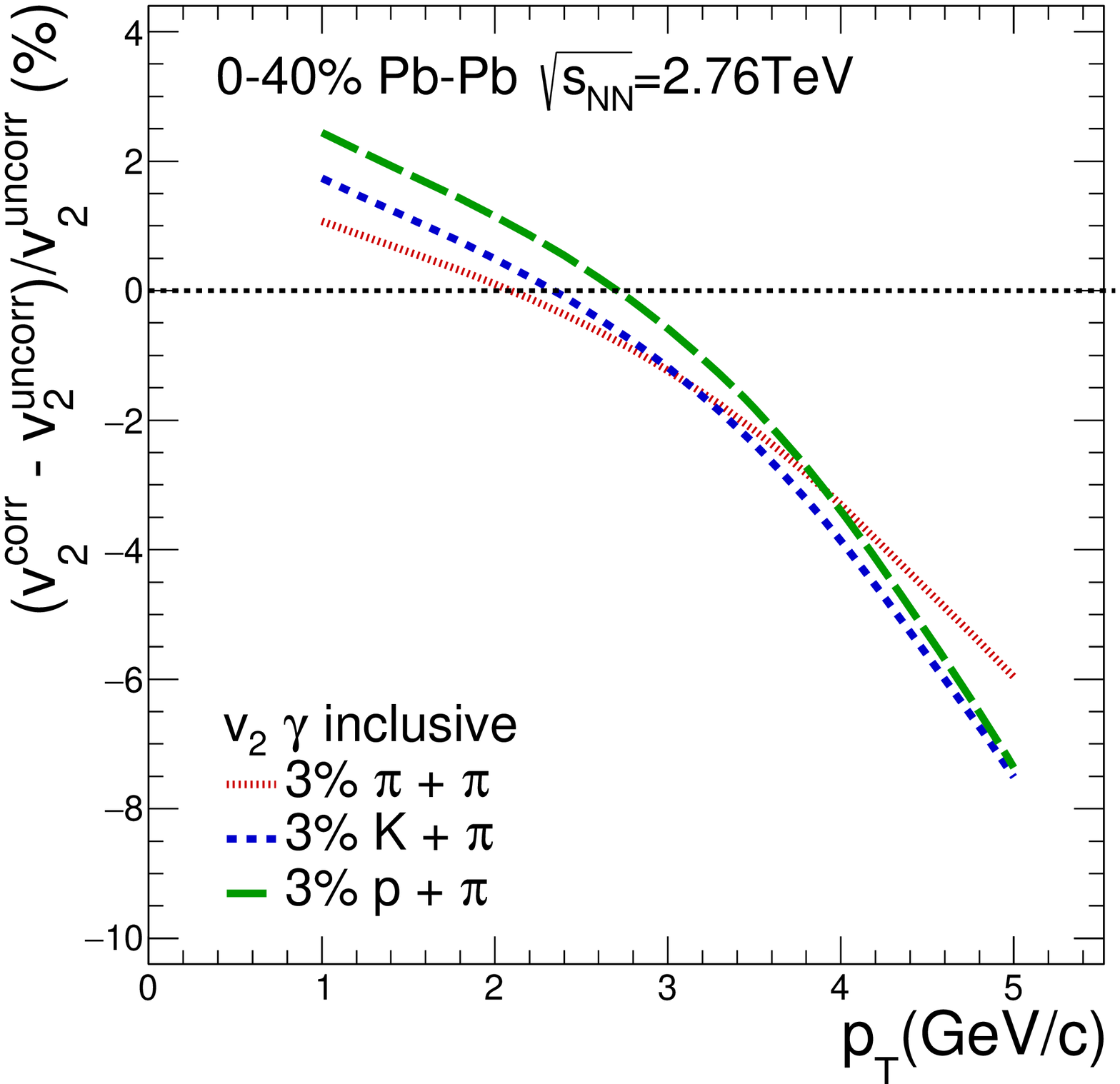}\\
\includegraphics[width=0.43\linewidth]{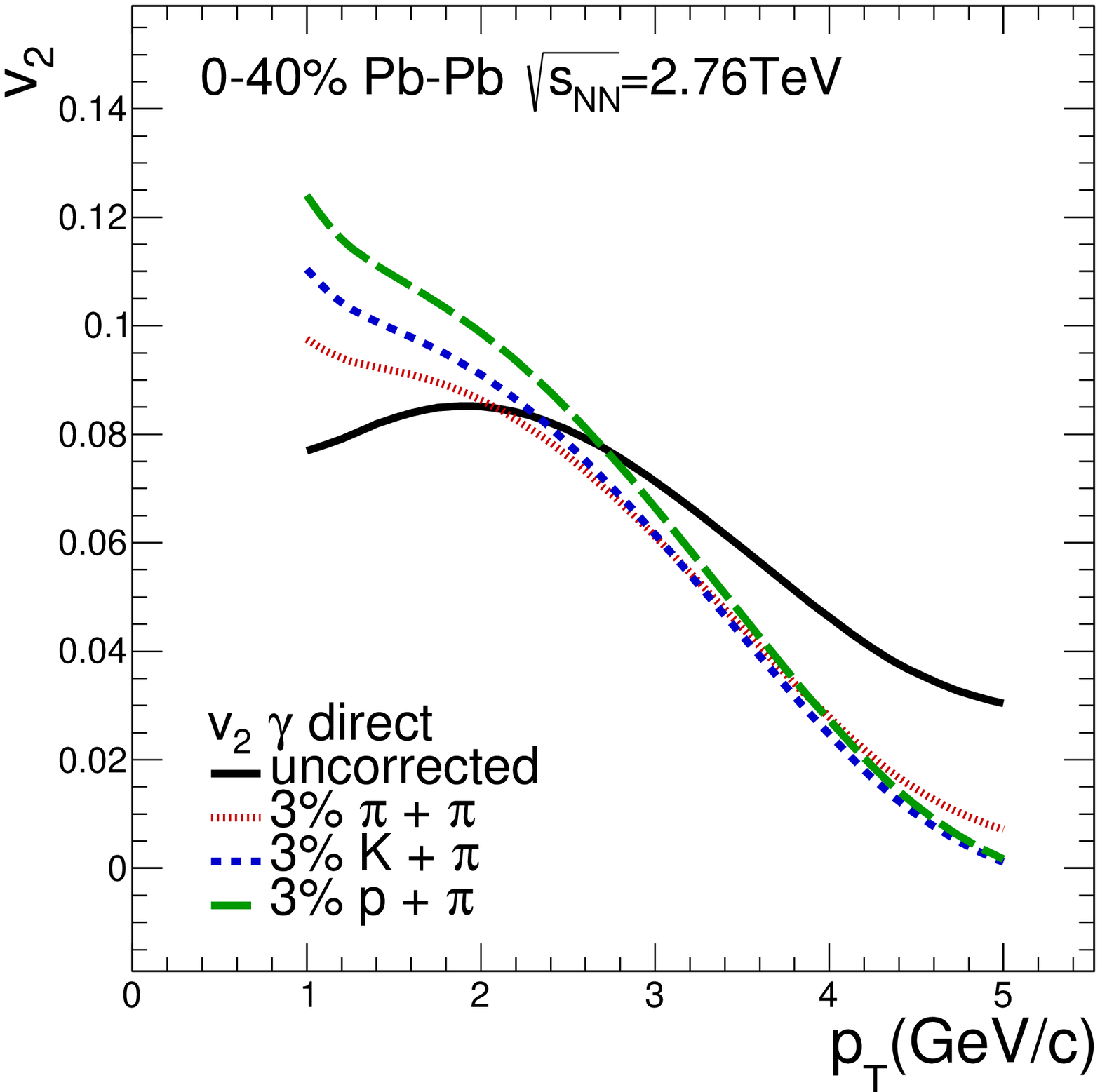}\hspace{0.5cm}
\includegraphics[width=0.43\linewidth]{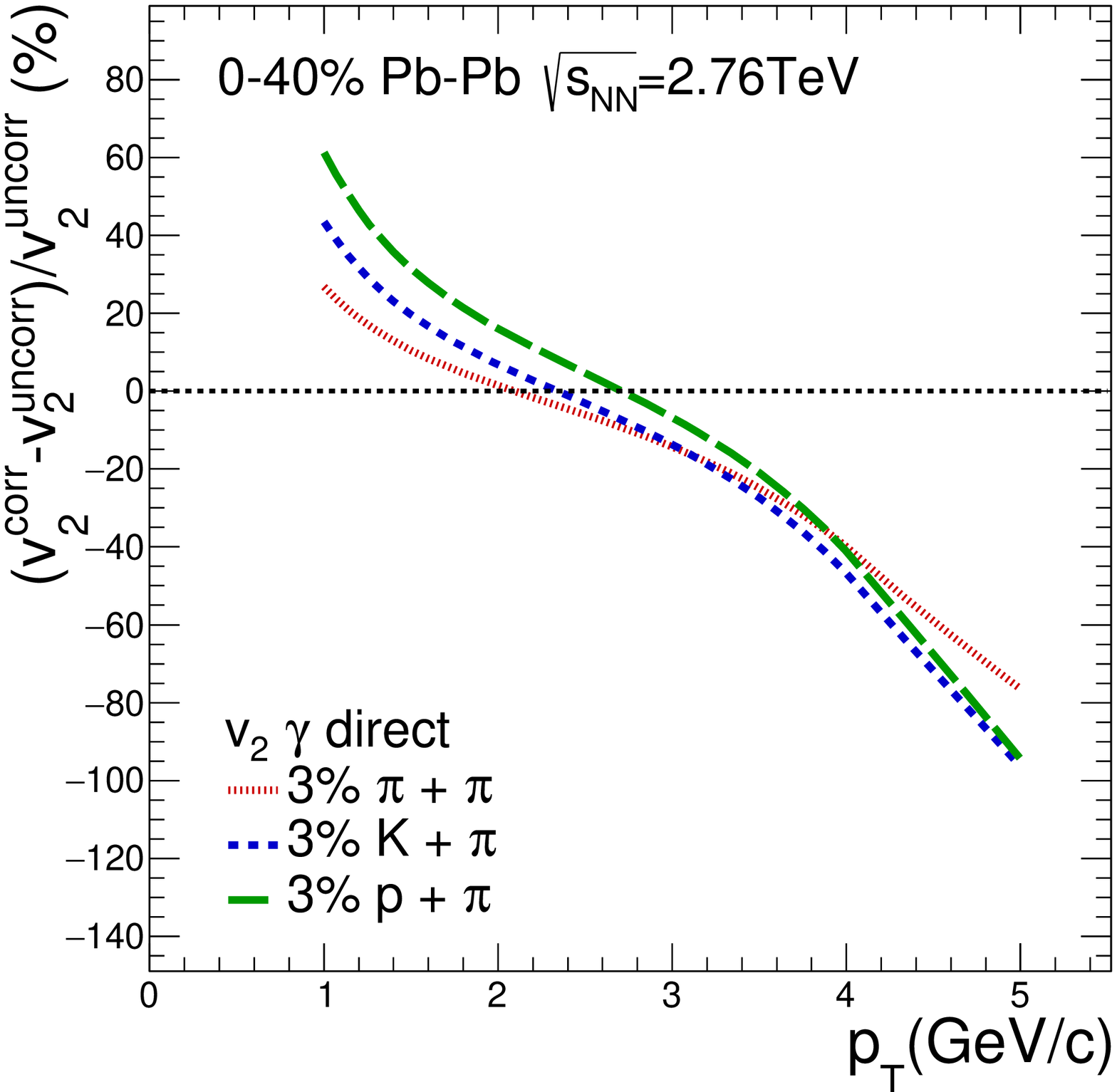}\\
\caption{Results of $\vincg$ (top left panel) and $\vdirg$ (bottom left panel) using \Eq{v2correction} to correct for $\pi + \pi$, $K + \pi$ and $p + \pi$ contamination with $c=3\%$. The deviations (in \%) from the uncorrected $\vincg$ and $\vdirg$ are shown in the corresponding panels on the right.}
\label{fig:v2Comparison}
\end{figure}

\begin{figure}[t!]
\centering
\includegraphics[width=0.43\linewidth]{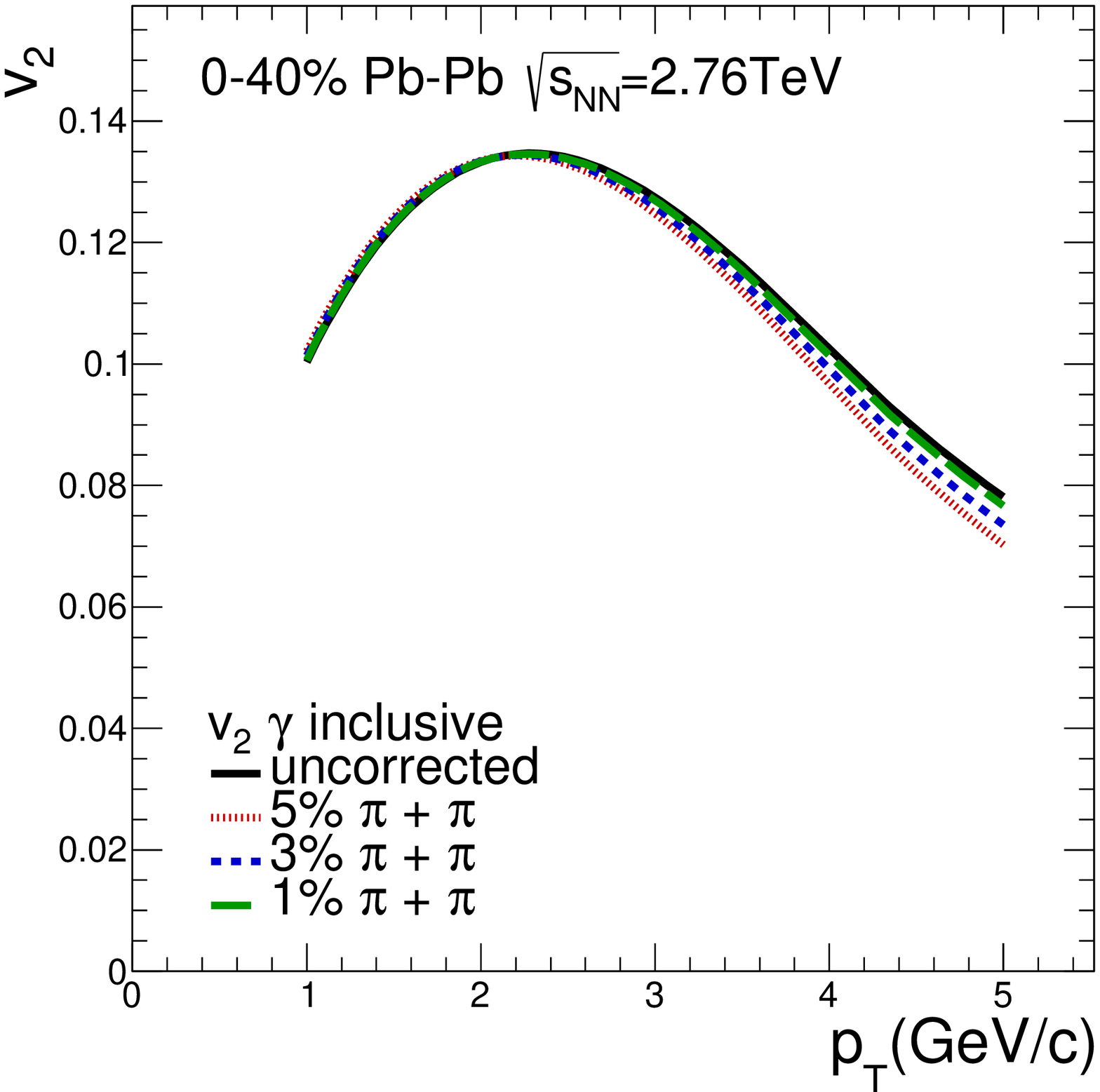}\hspace{0.5cm}
\includegraphics[width=0.43\linewidth]{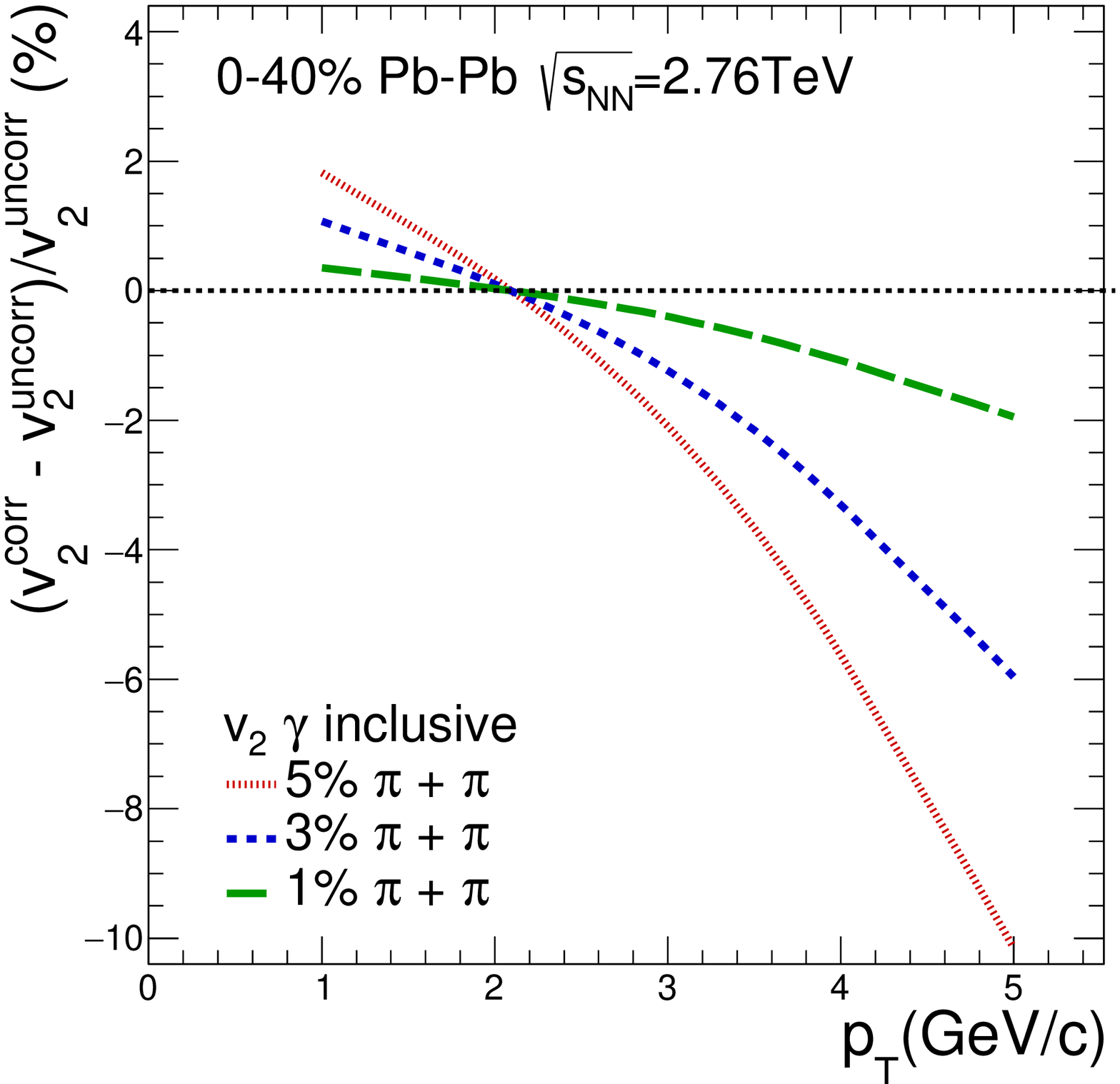}\\
\includegraphics[width=0.43\linewidth]{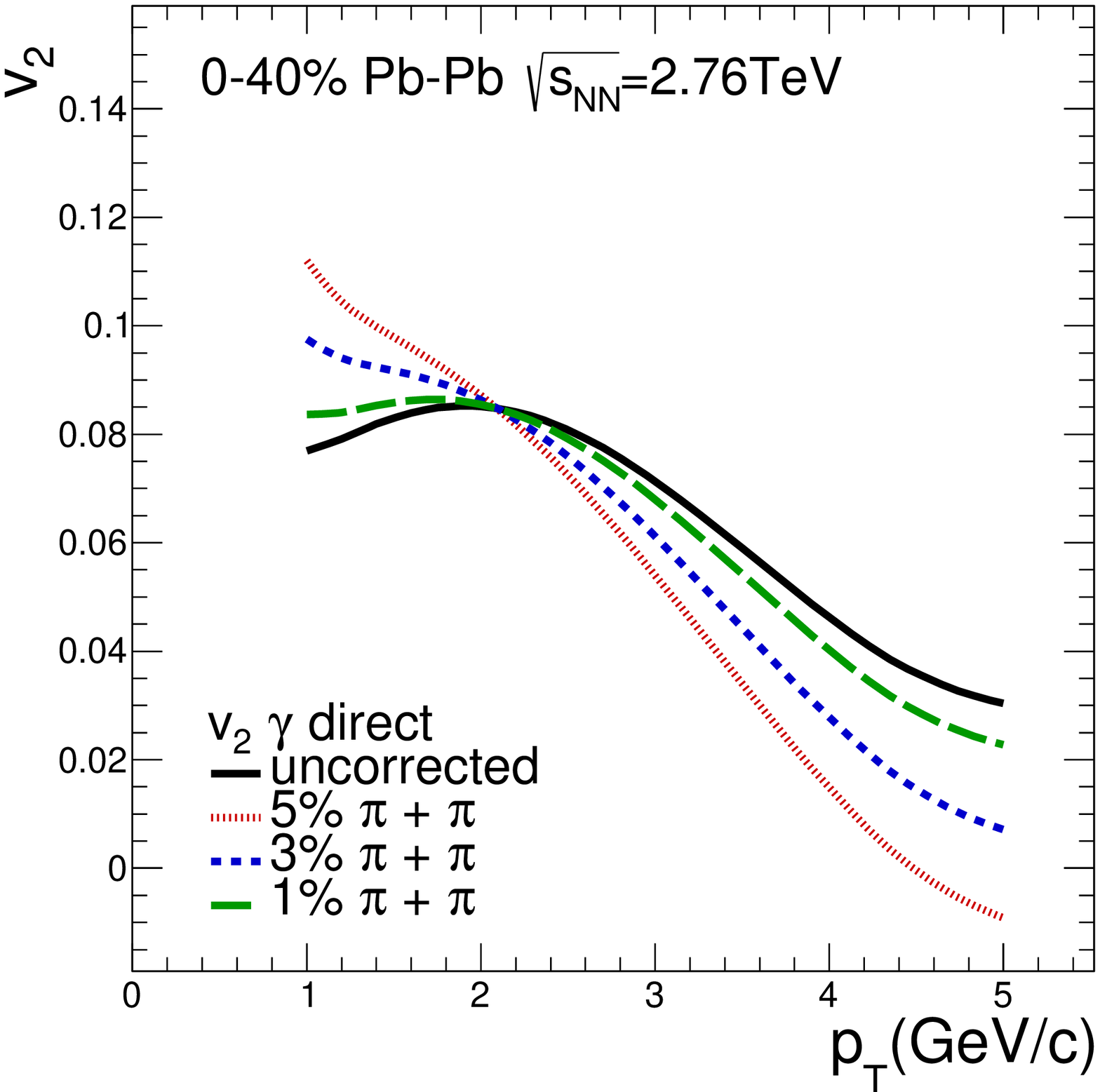}\hspace{0.5cm}
\includegraphics[width=0.43\linewidth]{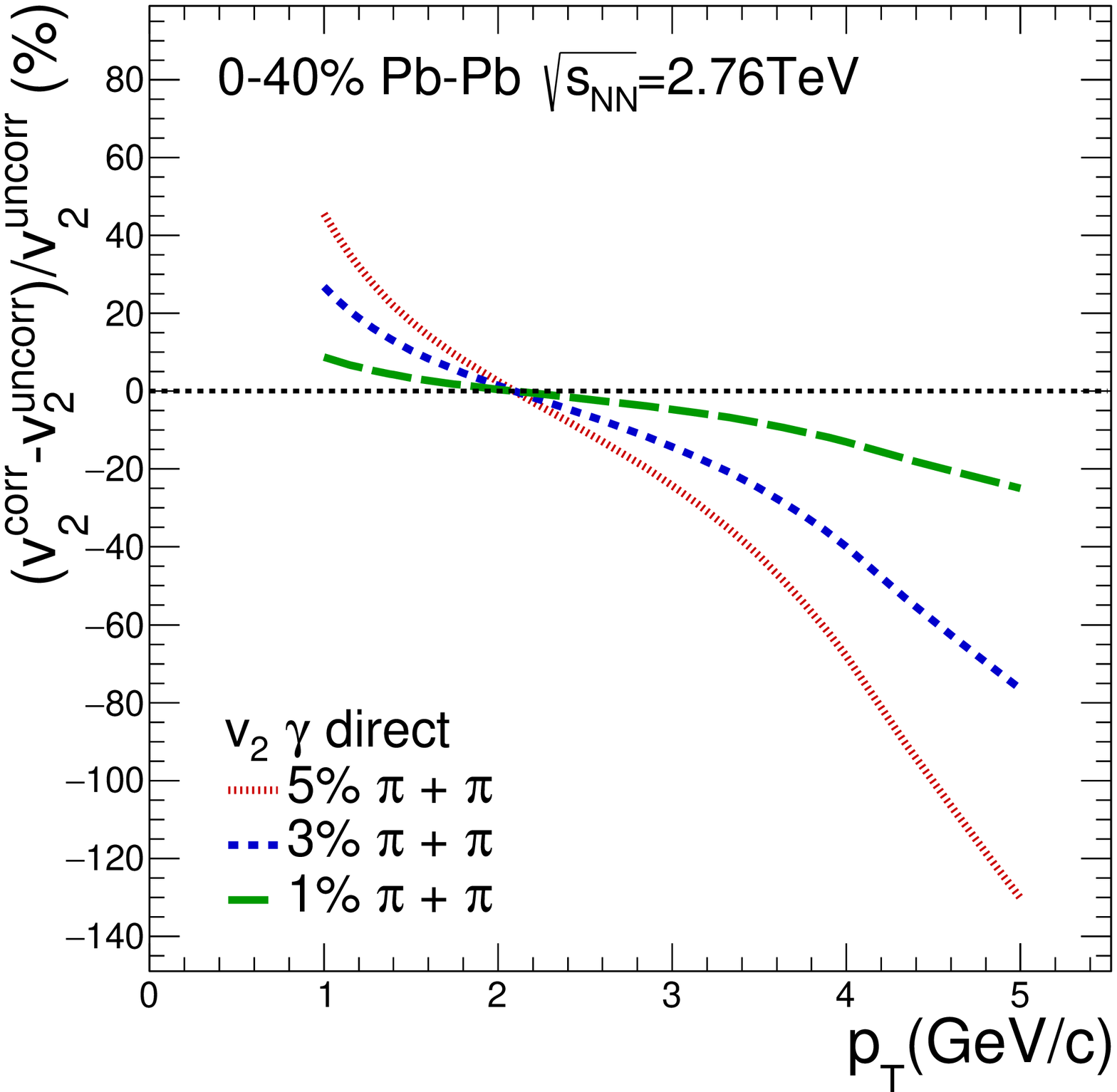}\\
\caption{Results of $\vincg$ (top left panel) and $\vdirg$ (bottom left panel) using \Eq{v2correction} to correct for $\pi + \pi$ with $c=1$, $3$ and $5$\%. The deviations (in \%) from the uncorrected $\vincg$ and $\vdirg$ are shown in the corresponding panels on the right.}
\label{fig:v2ComparisonPions}
\end{figure}

\section{Results}
\label{sec:res}
In this section, the inclusive photon flow $\vincg$ shown in \Fig{fig:ALICEInput} is corrected using \Eq{v2correction} for different assumptions of $\vbkgg$, shown in \Fig{fig:v2backgroundflow}, and purity.
The direct photon flow $\vdirg$ is calculated for the uncorrected and background corrected $\vincg$ using \Eq{v2directform}.

\Fig{fig:v2Comparison} illustrates the effect of background $v_2$ corrections on inclusive and direct photon flow for different assumptions on the type of background. 
A $\pT$ independent inclusive photon sample purity of 97\% is assumed, i.e.\ a contamination of $c=3$\% originating from $\pion+\pion$, $\kaon+\pion$ and $\prot+\pion$ pair $v_2$, respectively, is considered. 
As may be expected, the effects on inclusive photon $v_2$ are moderate, between $+2$\% and $-8$\%. 
The effect on the direct photon flow, however, is considerable, between $+60$\% and up to $-90$\% depending on $\pT$ and the type of background.
The differences for the different particle species contributing to the background are again rather moderate, obviously because the differences between the assumed pair $v_2$ are also rather small. 

A straightforward next step is the study of the dependence of the correction on the strength of the contamination, as shown in \Fig{fig:v2ComparisonPions}.
Here, only the shape of the $\pion+\pion$ pair $v_2$ is used as $\vbkgg$, but different levels of contamination of $c=1$, $3$ and $5$\%, respectively, are assumed. 
This leads to slightly stronger effects, in particular for $c = 5$\%. 
For inclusive photon $v_2$ effects are between $+2$\% and $-10$\%, for direct photons between $+50$\% and up to $-120$\%.
For $c=3$\% and restricting to about $3$~\GeVc\ the effect on $\vdirg$ is about $50$\%.

In our calculations the shift from the correction is positive at low $\pT$ and becomes negative at high $\pT$. 
We do not want to attribute too much significance to these features, as they depend crucially on details of the pair $v_2$ values used for correction, and the systematic uncertainty on those estimates is considerable. 
For example, if in reality the pair $v_2$ would be similar to the estimate of coalescence from data, shown in \Fig{fig:v2backgroundflow}, then the correction would imply a reduction of the $\vincg$, and hence the $\vdirg$, everywhere.
Precise estimates require to measure the pair $v_2$ in data, which is beyond the scope of this article.

\section{Summary}
\label{sec:sum}
ALICE preliminary $v_2$ results~(\Fig{fig:ALICEInput}) for inclusive and direct photons reconstructed from conversion electron and positron pairs are used to study the effect of a possible contamination of the inclusive photon sample.
The event generator Therminator2, which employs (2+1)-dimensional boost-invariant hydrodynamics to describe the single-particle $v_2$ coefficients~(\Fig{fig:v2singleflow}), is used to model the elliptic flow of the possible pair background in the photon conversion sample~(\Fig{fig:v2backgroundflow}). 
The effect on inclusive and direct photon $v_2$ of 3\% contamination from $\pion+\pion$, $\kaon+\pion$ and $\prot+\pion$ pairs~(\Fig{fig:v2Comparison}) and for varying contamination from $\pion+\pion$ pairs~(\Fig{fig:v2ComparisonPions}) has been studied.
While the effect on the inclusive photon $v_2$ is moderate, between $+2$\% to $-10$\%, the effect on the direct photon flow is considerable, between $+40$\% and up to $-120$\% depending on $c$ and $\pT$.
For $c=3$\% and restricting to about $3$~\GeVc\ the effect on $\vdirg$ is about $50$\%.
Below $2$~\GeVc\ the considered shape of $\vbkgg$ did not lead to a decrease of the direct $v_2$, however the correlated background originating from pairs with at least one e$^{\pm}$ has not been simulated.
Our calculations demonstrate that it is important to correct the inclusive photon sample as precisely as possible for even small impurities.

\section*{Acknowledgements}
The work of F.\ Bock and C.\ Loizides is supported in part by the U.S. Department of Energy, Office of Science, Office of Nuclear Physics, under contract number DE-AC02-05CH11231.
The work of M.\ Sas and T.\ Peitzmann is supported in part by the Stichting voor Fundamenteel Onderzoek der Materie (FOM), Netherlands.

\bibliographystyle{utphys}
\bibliography{biblio}
\end{document}